\documentclass[%
superscriptaddress,
reprint,
showpacs,preprintnumbers,
 amsmath,amssymb,
 aps,
pra,
]{revtex4-1}

\usepackage{graphicx}
\usepackage{dcolumn}
\usepackage{bm}
\usepackage{natbib}
\usepackage{gensymb}

\graphicspath{{images/}}
\begin{document}

\preprint{APS/123-QED}

\title{How confinement induced structures alter the contribution of hydrodynamic and short ranged repulsion forces to the viscosity of colloidal suspensions}

\author{Meera Ramaswamy}
\thanks{To whom correspondence should be addressed. Email: mr944@cornell.edu}
\affiliation{Department of Physics, Cornell University}
\author{Neil Y.C. Lin}
\affiliation{Department of Physics, Cornell University}
\author{Brian D. Leahy}
\affiliation{Department of Physics, Cornell University}
\author{Christopher Ness}
\affiliation{Department of Chemical Engineering and Biotechnology, University of Cambridge}
\author{Andrew Fiore}
\affiliation{Department of Chemical Engineering, Massachusetts Institute of Technology}
\author{James W. Swan}
\affiliation{Department of Chemical Engineering, Massachusetts Institute of Technology}
\author{Itai Cohen}
\affiliation{Department of Physics, Cornell University}
 
\date{\today}

\begin{abstract}
Confined systems ranging from the atomic to the granular are ubiquitous in nature. Experiments and simulations of such atomic and granular systems have shown a complex relationship between the microstructural arrangements under confinement, the short ranged particle stresses, and flow fields. Understanding the same correlation between structure and rheology in the colloidal regime is important due to the significance of such suspensions in industrial applications. Moreover, colloidal suspensions exhibit a wide range of structures under confinement that could considerably modify such force balances and the resulting viscosity. Here, we use a combination of experiments and simulations to elucidate how confinement induced structures alter the relative contributions of hydrodynamic and short range repulsive forces to produce up to a tenfold change in the viscosity. In the experiments we use a custom built confocal rheoscope to image the particle configurations of a colloidal suspension while simultaneously measuring its stress response. We find that as the gap decreases below 15 particle diameters, the viscosity first decreases from its bulk value, shows fluctuations with the gap and then sharply increases for gaps below three particle diameters. These trends in the viscosity are shown to strongly correlate with the suspension microstructure. Further, we compare our experimental results to those from two different simulations techniques, which enables us to determine the relative contributions of hydrodynamic and short range repulsive stresses to the suspension rheology. The first method uses the lubrication approximation to find the hydrodynamic stress and includes a short range repulsive force between the particles while the second is a Stokesian dynamics simulation that calculates the full hydrodynamic stress in the suspension. We find that the decrease in the viscosity at moderate confinements has a significant contribution from both the hydrodynamic and short range repulsive forces whereas the increase in viscosities at gaps less than three particle diameters arises primarily from short range repulsive forces. These results provide important insights to the rheological behavior of confined suspensions and further enable us to tune the viscosity of confined suspensions by changing properties such as the gap, polydispersity, and the volume fraction.
\end{abstract}

\maketitle


\section{Introduction}
Imagine driving on Delhi's narrow roads. The density of motorists is very high with vehicles ranging from large buses and trucks to motorbikes and auto-rickshaws all swerving in and out of their lanes. A similar drama unfolds in the confined flows of materials ranging from granular suspensions to colloids and even atoms. Determining how polydispersity, ordering, and confinement alter these flows in dense colloidal suspensions is particularly important since they are used extensively in industrial applications \cite{Oil, MotorOil}, automobile components \cite{fan2014polysulfide}, and common household products \cite{Food, Paint, cornstarch, shampoo}. Moreover, such suspensions display rheological properties that may usefully be compared to atomic systems at low shear rates \cite{PoonAtomCompare} as well as granular materials at high shear rates \cite{isa2007shear} and a microstructure that can be imaged in 3D using confocal microscopy. 

The ability to image the microstructure enables us to correlate the suspension structure with its rheology.  We focus on the dense suspension regime since in the dilute limit the detailed microstructure has little effect on the total shear viscosity \cite{Xiang}. At large volume fractions, however, many studies have shown a correlation between the microstructure and the rheology. For example, neutron scattering studies have shown that variations in the viscosity can be observed when structural changes occur in colloidal crystals \cite{Chen1, Chen2}. Further, numerical and theoretical calculations in colloidal crystals have indicated that the high frequency viscosity depends on the crystal structure and packing \cite{CalcViscCrystal1, CalcViscCrystal2}. 

The regime of confined flows is especially interesting since suspensions often display a rich range of structures below gaps of $\sim$10 particle diameters. For instance, free energy calculations show the existence of over 20 distinct crystalline arrangements when colloidal spheres are confined in gaps ranging from one to five particle diameters \cite{Struc7, khadilkar2016phase}. Many of these structures have also been observed experimentally \cite{Pieranski, ArjunYodh, Struc8, riley2010confinement, avendano2013directed}. Under shear, these arrangements often align with the direction of flow \cite{Cohen, XiangShear}. The vast range of structures formed under confinement \cite{Rice} suggests dense colloidal suspensions may have a rich variation in their shear viscosity. 

Further motivation to study the rheology of confined colloidal suspensions comes from granular systems, where experiments and simulations demonstrate a variety of viscosity trends under confinement \cite{PeylaExp, PeylaSim, Brown}.  At low volume fractions, experiments show a decrease in viscosity with decreasing gap followed by an increase in the viscosity at gaps corresponding to less than a few particle diameters \cite{PeylaExp}. In contrast, larger volume fractions show no overall decrease in the viscosity. Instead, fluctuations are observed that correlate with the incommensurability of the gap with the particle diameter \cite{Brown}. However, the large particle sizes in granular suspensions makes it difficult to image and hence correlate the microstructure and the rheology. In addition, very little is known about the origin of these changes in the rheology. Some experiments attribute the viscosity increase at extreme confinement to hydrodynamic forces \cite{PeylaExp} while other studies suggest that friction is responsible \cite{Brown}. This murkiness arises in part due to the difficulty of conducting studies that combine measurements of structure and rheology with simulations in order to distinguish how different structures alter the relative contributions of hydrodynamic and short ranged interaction forces. 

Such studies in granular suspensions suggest a similarly rich interplay will occur in colloidal systems. However, while there have been extensive investigations of the many structural transitions for colloidal suspensions under confinement, the rheological properties for systems with small gaps are poorly understood, in part due to lack of appropriate instrumentation. In particular, cone and plate rheometers have a varying gap across the shear region, and Couette rheometers have a fixed gap, which is difficult to control with micron scale precision. Parallel plate rheometers with circular flow can achieve small gaps but have a radially varying shear rate. Moreover, these rheometers are seldom coupled to microscopes making it difficult to correlate the microstructure and the rheology. Early attempts to simultaneously image the particle arrangements under confined flows studied suspension transport through capillaries and showed the flow rate changes with the particle density and ordering \cite{Seshadri, Poon}. In such measurements, however, it is difficult to determine a structure dependent viscosity since the total flow rate results from an average over a range of shear rates.  
	
These limitations can be overcome in simulations of confined suspension flows where several studies have shown that hydrodynamic lubrication forces alone can cause an increase in the viscosity of a suspension confined between two parallel walls \cite{Swan2, Bhattacharya1}. Other studies have demonstrated oscillations in viscosity and in normal forces \cite{gallier2016effect} similar to the fluctuations seen in granular systems \citep{Brown}. Such studies, however, are seldom compared to experiments. Without such comparisons, it is difficult to rule out contributions due to Brownian interactions and short ranged repulsion, that are thought to play a role at low and high shear rates respectively \cite{Xiang, Brady, wyart2014discontinuous, seto2013discontinuous}. 

	Here, we use a custom built parallel plate shear cell with translational flow that loads onto a confocal microscope to correlate the confinement induced microstructure with the confined suspension rheology. Further, we compare the experimentally measured viscosities to those from the lubrication-repulsion dynamics and Stokesian dynamics simulations. This comparison enables us to determine how microstructure alters the balance of short ranged and hydrodynamic forces to determine the measured viscosity trends.

\section{Apparatus and Methods}

\subsection{Experiments}

\begin{figure*}
\begin{center}
 \includegraphics[scale=0.55]{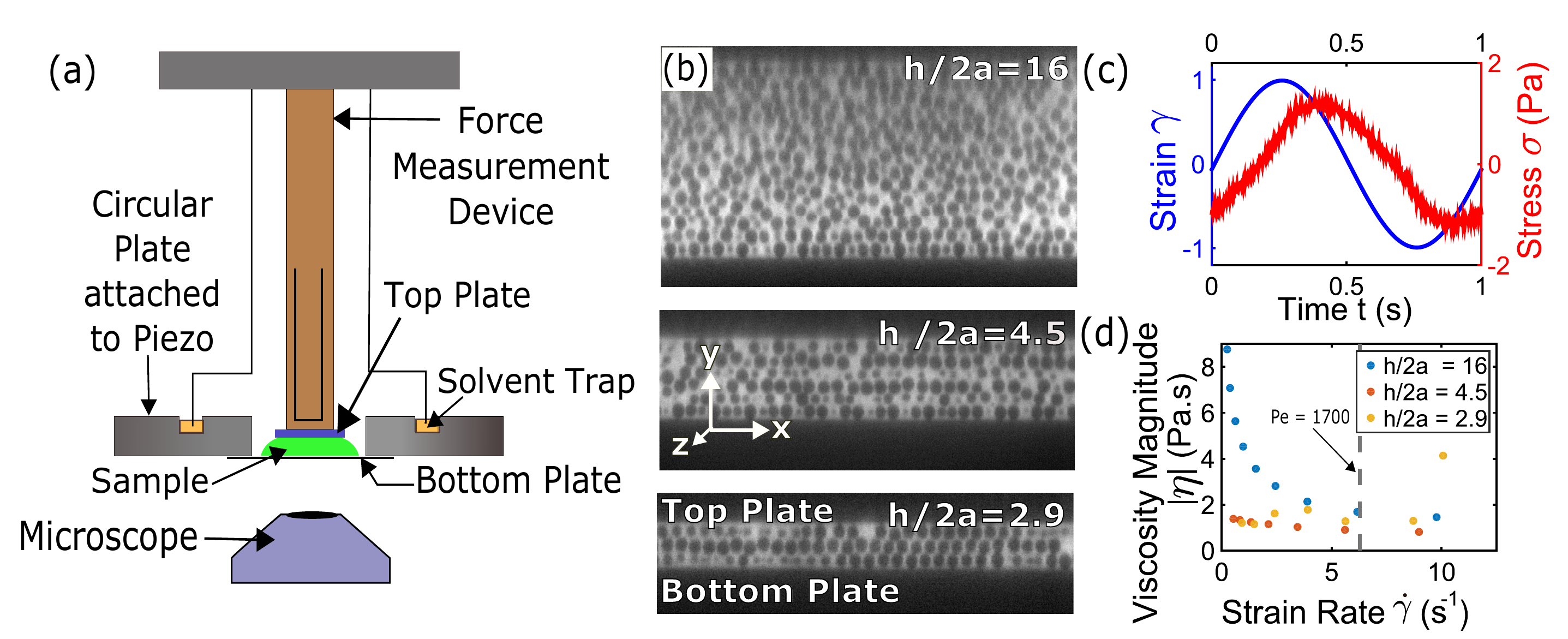}
\end{center}

\caption{The experimental apparatus and sample measurements. (a) A schematic of the shear cell, focusing on the shear zone and the force measurement device. (b) Confocal images in the shear gradient plane at three different extents of confinement. (c) The applied strain generated by the piezoelectric stage is depicted by the blue curve. A typical stress measurement that is obtained from the force measurement device is depicted by the red curve.  Importantly, to extract the viscosity magnitude we use imaging to back out the effective strain amplitude. (d) The magnitude of the complex viscosity as a function of the effective strain rate for the three gaps shown in (b). The vertical gray dashed line indicates the effective strain rate that is used for the remainder of the experiments.}
\label{fig:apparatus}
\end{figure*}

 To study the effect of confinement induced structures on the suspension rheology, we use a custom built confocal rheoscope \cite{LinShearCell}. A schematic of the device is shown in Fig.~\ref{fig:apparatus}a. Briefly, the shear cell has a bottom plate that is a transparent glass cover slip. The plate is attached to a piezoelectric stage that can translate along the flow, x, and gradient, y directions. The top boundary of the shear cell is a $16mm^{2}$ silicon wafer, which is atomically flat. The wafer is glued with epoxy to a force measurement device. The shear zone is surrounded by a suspension reservoir that maintains a constant osmotic pressure boundary condition (Fig.~\ref{fig:apparatus}a). A solvent trap is used to prevent evaporation of the suspending fluid. To achieve a uniform gap, the bottom plate and the force measurement device are attached to mounting brackets that can be adjusted using three set screws. The plates can be made parallel to within $4.3 \times 10^{-3}$ degrees with a gap h that can reach $2\mu m$. The device enables us to simultaneously shear the suspension, measure its rheology, and image its structure over a range of gaps as shown in Fig.~\ref{fig:apparatus}.   

The suspension consists of $2\mu$m diameter silica micropearl particles from Sekisui Chemical Company. The particles are suspended in a refractive index matching mixture of glycerol and water, that is 80-20 by mass fraction of glycerol water ($\eta_0 = 0.06$ Pa.s at $20 \degree$C). A small amount of fluorescein dye (2 mg/mL) is added to the solvent to enable imaging. The volume fraction of silica in the suspension is $0.52$, the densest suspension we could load and confine in our apparatus. For denser suspensions, the confining forces while loading the top plate are too large and the bottom glass plate breaks. This volume fraction corresponds to the crystal gas coexistence regime in hard spheres. The sample is sonicated and degassed to remove air bubbles prior to loading into the shear cell. 

A linear oscillatory strain is applied to the sample using the piezoelectric stage attached to the bottom plate, while nanometer scale deflections of the top plate are used to determine the force transmitted through the suspension. Examples of typical stress and strain curves are shown in Fig.~\ref{fig:apparatus}c. The system response is largely linear, with a measured stress that is nearly sinusoidal. From the Fourier transform of the stress response, we find that the third harmonic is smaller by at least factor of 10 for normalized gaps greater than 3. Therefore we report the magnitude of the complex viscosity associated with the first harmonic of the applied frequency.

Gap uniformity in this device is a major challenge since slight deviations from parallel alignment can generate unintended parasitic flows. Such effects are particularly prominent at small gaps where slight variations can lead to large changes in the applied strain. Thus, to set the gap between the plates of the shear cell, we use a painstaking imaging procedure. Briefly, we use the confocal microscope to image the entire gap at 9 equally spaced locations within the shear zone. We analyze these images and determine the derivative or change in the total intensity as a function of height y. The distance between the maximum and the minimum of this curve gives the gap between the plates to a precision of $0.1$ microns.	Importantly, the suspension is relaxed overnight so that any stress bowing the bottom cover slip dissipates. To confirm that this method is precise and that the force measurement device works accurately at small gaps, we measure the viscosity of a Newtonian fluid. We find that the fluid viscosity is constant over the range of gaps (2$\mu$m - 100$\mu$m) in which we are interested, confirming our excellent control over the shear geometry.
	
Wall slip presents an additional challenge in dense suspensions. Typical methods to prevent wall slip such as roughening the boundaries of the shear zone can no longer be used as they will cause complications during gap alignment and imaging. Instead, we measure the effective strain in the system by imaging the top and the bottom particle layers. Particle Image Velocimetry is used to find the average displacements of the particles at the boundaries. The difference between the displacement of the bottom and top layers is divided by the distance between the centers of the particles to calculate the effective strain in the sample. Importantly, in order to compare the suspension response at different gaps, we had to conduct preliminary strain sweeps at each gap to determine the { \it applied strain} that generates the desired {\it effective strain}. 

For shear experiments on suspensions, the measured force response is a sum of three major contributions: the Brownian, hydrodynamic and short-ranged repulsive stresses \cite{Lin, wyart2014discontinuous, seto2013discontinuous, Brady, sekimoto1993mechanism, maranzano2001effects}. The relative magnitude of the Brownian and hydrodynamic interactions is characterized by the Peclet number for the system, $Pe = 6 \pi \eta \dot{\gamma} a^{3} / k_{B} T $ where $\eta$  is the viscosity of the fluid, $\dot{\gamma}$ is the shear rate, $a$ is the radius of the particle, $k_{B}$ is the Boltzmann constant and $T$ is the temperature. Since simulations of colloidal and granular systems have both implicated hydrodynamic lubrication forces as the origin for the rapid viscosity increase as the gap is decreased to several particle diameters \cite{PeylaSim,Swan1}, we focus on the large $Pe$ regime where Brownian forces are negligible, the system is no longer thinning, and the viscosity depends weakly on the shear amplitude (\cite{Xiang, PNASThinning}). For the measurements reported here, we achieve a dominantly linear stress-strain response with $Pe \approx 1700$ for all gaps by shearing at a frequency of 1Hz and an effective strain amplitude of $\gamma_0 = 1$ as shown in Fig.~\ref{fig:apparatus}d. Moreover, due to the small particle size, the Reynolds number is extremely small and particle inertia can be neglected. 

At these large Peclet numbers $\approx 1700$, it has been suggested that short range interparticle repulsive forces contribute to the suspension stress \cite{sierou2002rheology, melrose2004continuous}. These repulsive forces can arise from various sources ranging from actual contact to screened electrostatic repulsion between the particles. Here, we remain agnostic to the origin of these forces and use the term \emph{short range repulsion} to refer to them collectively.

\subsection{Simulations}
To develop an understanding of the different contributions to the stress characterizing the suspension rheology, we conduct two different simulations. The first is a lubrication-repulsion model that approximates the hydrodynamic stresses using a lubrication approximation between particle pairs and introduces a steep repulsive particle-particle interaction to prevent overlaps. The second is a Stokesian Dynamics Simulation that includes both the short range and the long range contribution to hydrodynamic forces. Comparison between these models and the data allow for determining: 1) the fidelity of the calculations to the experimental measurements 2)  whether the lubrication approximation accurately accounts for the full hydrodynamic stresses under confinement, and 3) whether hydrodynamic interactions alone account for the experimentally measured changes in the viscosity. 

\subsubsection{Lubrication-repulsion Dynamics}

In the lubrication-repulsion model, we solve the equations of motion for non-inertial, non-Brownian spheres of diameter $2a_{i}$ translating and rotating with velocity vectors $\mathbf{v}$ and $\boldsymbol{\omega}$ respectively. The spheres are suspended in a density-matched fluid of viscosity $\eta_f$. The particles are subjected to forces arising due to hydrodynamics and repulsive particle-particles interactions. For efficient computation, we neglect all hydrodynamic terms other than the divergent, short-range, pairwise lubrication forces between neighboring particles~\cite{Kim1991}. Briefly, this approximation is valid because in the near field limit, the forces diverge as $1/h$ and torques diverge as $\log 1/h$ where $h$ is the ratio of the distance between the particle surfaces to the particle diameter $2a$. However, the many-body  and the far field terms fall of as $1/r$ where r is the distance between the centers of the spheres. In the limit $h\ll a$, the two body resistance terms dominate over the many-body and far field terms ~\cite{Ball1997,Kumar2010, Trulsson2012,Mari2015,Ness2016a} and this approximation has shown to deliver useful quantitative results for dense suspensions where the volume fraction $\phi$ is $\gtrsim 0.4$.

The lubrication-repulsion model calculates the hydrodynamic force and torque on particles $i$ due to particle $j$, with $\mathbf{r}_{ij}$ the vector pointing from $j$ to $i$ and $\mathbf{n}_{ij}=\mathbf{r}_{ij}/|\mathbf{r}_{ij}|$, the force $\mathbf{F}^{h}_{ij}$ and torque $\bm{\Gamma}^{h}_{ij}$ as
\begin{widetext}
\begin{subequations}
\begin{align}
&\mathbf{F}^{h}_{ij} = -a_{sq} 6 \pi \eta_f (\textbf{v}_i - \textbf{v}_j) \cdot \textbf{n}_{ij} \textbf{n}_{ij}- a_{sh} 6 \pi \eta_f (\textbf{v}_i - \textbf{v}_j) \cdot (\mathbf{I}-\mathbf{n}_{ij}\mathbf{n}_{ij}) \text{,}\\
&\bm{\Gamma}^{h}_{ij} = -a_{pu} \pi \eta_f (2a_i)^3(\boldsymbol{\omega}_i - \boldsymbol{\omega}_j) \cdot (\mathbf{I}-\mathbf{n}_{ij}\mathbf{n}_{ij}) - a_i \left(\mathbf{n}_{ij} \times \mathbf{F}^{h}_{ij}\right) \text{,}
\end{align}
\label{eq:lube_forces}
\end{subequations}
\end{widetext}
for $3\times3$ identity tensor $\mathbf{I}$ and squeeze $a_{sq}$, shear $a_{sh}$ and pump $a_{pu}$ resistance terms~\cite{Kim1991}, with $\beta = a_j/a_i$, as

\begin{widetext}
\begin{subequations}
\begin{align}
&a_{sq} = \frac{2 \beta^2}{(1+\beta)^2} \frac{a_i^2}{h_\text{eff}}	+	\frac{1 + 7\beta + \beta^2}{5(1 + \beta)^3} a_i\ln \left(\frac{a_i}{h_\text{eff}} \right) + \frac{1 + 18\beta - 29\beta^2 + 18\beta^3 + \beta^4}{21(1+\beta)^4} \frac{a_i^2}{h_\text{eff}} \ln \left(\frac{a_i}{h_\text{eff}}\right) \text{,}\\
&a_{sh} = 4 \beta \frac{2 + \beta + 2\beta^2}{15 (1 + \beta)^3} a_i \ln \left( \frac{a_i}{h_\text{eff}}\right) + 4\frac{16 -45\beta + 58\beta^2 - 45\beta^3 + 16\beta^4}{375(1 + \beta)^4} \frac{a_i^2}{h_\text{eff}} \ln \left( \frac{a_i}{h_\text{eff}} \right) \text{,}\\
&
a_{pu} = \beta \frac{4 + \beta}{10(1 + \beta)^2} \ln \left( \frac{a_i}{h_\text{eff}} \right) + \frac{32 - 33\beta + 83\beta^2 + 43\beta^3}{250\beta^3} \frac{a_i}{h_\text{eff}} \ln \left( \frac{a_i}{h_\text{eff}} \right) \text{.}
\end{align}
\label{eq:resistances}
\end{subequations}
\end{widetext}

The surface-to-surface distance $h$ is calculated for each pairwise interaction according to $h = |\mathbf{r}_{ij}| - a_i + a_j$.
We truncate the lubrication divergence and regularize the contact singularity at a typical asperity length scale $h_\text{min}=0.002a_{ij}$, where $a_{ij} = \frac{a_ia_j}{a_i + a_j}$ is the weighted average particle radius. We set $ h = h_\text{min}$ in the hydrodynamic force calculation, when $h < h_\text{min}$.
The effective interparticle gap used in the force calculation, $h_\text{eff}$, is therefore given by
\begin{equation}
    h_\text{eff} =  \left\{
                \begin{array}{ll}
                  h & \text{for } h > h_\text{min}\\
                  h_\text{min} & \text{otherwise.} 
                \end{array}
              \right.
\end{equation}
For computational efficiency, the lubrication forces are omitted ($\mathbf{F}^h_{ij}$, $\mathbf{\Gamma}^h_{ij} = 0$) when the interparticle gap $h$ is greater than $h_\text{max} = 0.1 a_{ij}$. The volume fraction is sufficiently high in the present work that all particles have numerous neighbors with $h < h_\text{max}$, so such an omission is inconsequential to the dynamics.

The lubrication-repulsion model further applies a penalty function to minimize overlap between spheres for which $h<0$, representing a generic particle-particle repulsive potential. For simplicity, the interaction is modeled as a linear spring~\cite{Cundall1979}, with a normal repulsive force given by
\begin{equation}
\mathbf{F}^{c}_{ij} =  \left\{
                \begin{array}{ll}
                  k \delta \mathbf{n}_\text{ij} & \text{for } \delta > 0\\
                  0 & \text{otherwise,} 
                \end{array}
              \right.
\end{equation}
for spring stiffness $k$ and particle overlap $\delta$ equivalent to $-h$. We find that the simulation results do not depend sensitively on the value of k or on whether the contact is Hertzian or Hookean. In particular, increasing k over three orders of magnitude does not quantitatively change the results.

Hydrodynamic and short range repulsive forces are summed on each particle, and the trajectories are updated in a step-wise, deterministic manner according to a Velocity-Verlet scheme. The computational model is implemented in \texttt{LAMMPS}~\cite{Plimpton1995}.


To perform confinement simulations using the lubrication-repulsion model, a shear cell is constructed with upper and lower confining walls normal to y, with the separation between the walls being prescribed in advance to achieve a desired confinement. The walls measure $60a\times60a$ and are bound by periodic boundaries in x and z. The walls are constructed from dense arrays of fixed particles with diameters one-tenth that of the suspension particles. The walls interact with suspension particles through the above repulsive forces~\cite{Singh2000,Chow2015}. The gap between the shear cell walls is initially populated with randomly located particles that are allowed to relax before shearing commences. 

Taking the simulation cell as a representative control volume $V$, the corresponding $3\times3$ bulk stress tensor is calculated according to
\begin{equation}
\bm{\sigma} = \frac{1}{V} \left[\sum_i\sum_{j\neq i} \mathbf{r}_{ij} \mathbf{F}^h_{ij}+ \sum_i\sum_{j\neq i} \mathbf{r}_{ij} \mathbf{F}^c_{ij} \right]\text{.} 
\label{eq:NetStress}
\end{equation}
The shear stress of interest is the $\sigma_{xy}$ component of $\bm{\sigma}$. Samples are sheared at a strain amplitude of 1 and at a frequency that gives a characteristic Reynolds number of 0.01 producing over-damped dynamics such as those found in the experiments. For each simulation, the sample was sheared for 10 cycles and the stress from the final cycle is used to calculate the viscosity in a manner similar to the experiments. The quantities $\rho \dot{\gamma} d^2/\eta_f$ and $\dot{\gamma}d/\sqrt{k/\rho d}$ remain $\ll1$, ensuring non-inertial and \emph{nearly}-hard particle rheology throughout.

\subsubsection{Stokesian Dynamics}

Here, we use a variation of the Stokesian Dynamics algorithm to compute the total hydrodynamic contribution to the viscosity of the suspension. Since current simulation techniques using Stokesian Dynamics are extremely slow for systems with more than 1000 particles, we use Brownian Dynamics simulations to generate particle trajectories. These configurations generated with Brownian Dynamics are expected to be representative of those from standard Stokesian Dynamics because the particle volume fraction is large and the particles are strongly confined, so that hydrodynamic interactions are screened \cite{koumakis2016start}. From the particle trajectories, we compute the total hydrodynamic contribution to the viscosity using the Stokesian Dynamics approach.

In the Brownian Dynamics simulations we start by placing 2000 particles randomly in a large simulation box (volume fraction, $\phi=0.05$). The system is periodic in all three dimensions. The system is thermally equilibrated for 100 particle diffusion times while shrinking the box in the unconfined dimensions until a volume fraction $\phi=0.52$, which matches the experimental conditions, is reached. The particle trajectories during equilibration are generated by over-damped Brownian Dynamics simulations using the HOOMD-Blue software package \cite{Anderson, Glaser}. After the equilibration period, the system is sheared at a strain amplitude 1.3, and frequency 1 Hz for 100 cycles, with configurations output for analysis 10 times per cycle. The linear shear rate $\dot{\gamma}$ is implemented using the Lees-Edwards boundary condition \cite{lees1972computer}. The particles are represented in the simulation as hard spheres, with the hard sphere constraint implemented by the Heyes-Melrose algorithm \cite{heyes1993brownian}, which applies a pairwise spring-like conservative force to all overlapping particle pairs. A detailed description of the hard sphere constraints and Brownian Dynamics methodology used here is described elsewhere for the case of oscillatory shear \cite{wang2016large}. The impenetrable walls are implemented through the built-in HOOMD wall class using a purely repulsive shifted Lennard-Jones potential to represent the particle-wall interactions,

\begin{widetext}
\begin{equation}
		V\left(r\right) = 	\begin{cases}
								4 \epsilon \left[ \left(\frac{\sigma}{r}\right)^{12} - \alpha \, \left(\frac{\sigma}{r}\right)^{6} \right] - \left(r-r_{\rm cut}\right)\frac{\partial V_{\rm LJ}}{\partial r}\left(r_{\rm cut}\right), & r < r_{\rm cut} \\
								0, & r \geq r_{\rm cut}
							\end{cases}	
\end{equation}
\end{widetext}
	where $V_{\rm LJ}$ is the standard Lennard-Jones potential. For the purely repulsive wall potential, the dimensionless simulation parameters are, $\alpha=0$, $\sigma=1$, $\epsilon=1$, where distance $\sigma$ is made dimensionless on the particle radius $a$, and energy $\epsilon$ is made dimensionless on the thermal energy multiplied by the Peclet number, $k_{B}T \, {\rm Pe}$, ${\rm Pe}=6\pi\eta\dot{\gamma}{a^{3}}/k_{B}T$. This energy scaling ensures that the wall forces are strong enough to prevent overlap in the sheared system. The particle-wall interactions are truncated at $r_{\rm cut}=a$ so that particles only experience wall forces when they overlap the wall. 
	
	The configurations along the trajectories generated with Brownian Dynamics are used to compute the hydrodynamic contribution to the viscosity via Stokesian Dynamics \cite{Swan2}. The specific quantity reported is the high frequency shear viscosity, calculated as the mean hydrodynamic stresslet for a particular configuration. The calculated viscosity is a sum of the long range hydrodynamic and short range lubrication contributions to the \emph{hydrodynamic} stress and does not include any short range repulsive or Brownian contributions to the stress.  
	
\section{Rheology}

\begin{figure}
\begin{center}
\includegraphics[scale= 0.5]{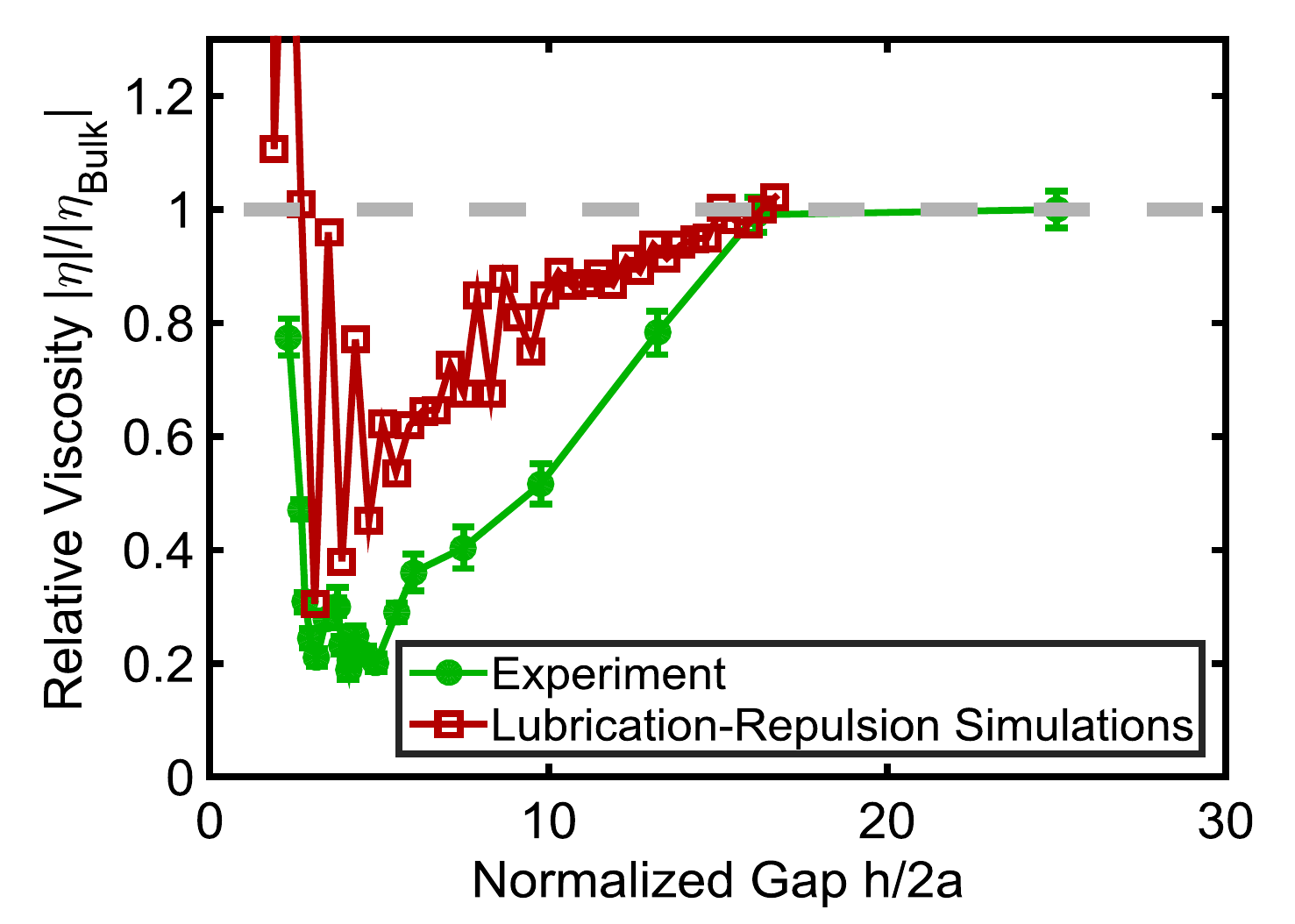}
\end{center}

\caption{Magnitude of the normalized complex viscosity versus normalized gap. A fivefold decrease in the viscosity is observed below gaps smaller than 15 particle diameters. For gaps lower than three particle diameters a steep increase in the viscosity is observed. Very similar trends are observed in the experiment (green circles) and simulation (red squares) data. Uncertainty in the experiments corresponds to the level of background noise. The gray dashed line indicates the viscosity trend for a Newtonian fluid.}
\label{fig:rheology}
\end{figure}

To determine how the suspension rheology is altered by confinement, we plot the magnitude of complex viscosity $\eta$ normalized by the bulk suspension viscosity $\eta_{Bulk}$ as a function the normalized gap, h/2a at an effective strain rate amplitude of $2 \pi$ s$^{-1}$ in Fig.~\ref{fig:rheology}. From the experiments, we find that under increasing confinement, changes in the viscosity from bulk can be broken up into three regimes discussed in greater detail below: 1) a decrease in the viscosity for 15 $ > $ h/2a $>$ 6, 2) smaller scale fluctuations in the viscosity when 6 $ >$ h/2a, and 3) a sharp increase in the viscosity when 3 $>$ h/2a. We find that the lubrication-repulsion dynamics simulations captures these trends (red symbols in Fig.~\ref{fig:rheology}). We use our imaging capability to test the hypothesis that changes in the suspension microstructure are correlated to the observed variations in viscosity in each of these regimes. To address whether these structural changes act through short range repulsive forces or hydrodynamics both of which are present in the lubrication-repulsion dynamics simulations, we compare our results to the Stokesian Dynamics simulations, which calculate only the hydrodynamic stress contributions and do not include additional short-ranged repulsive interactions. 
 
\subsection{Moderate Confinement}
\begin{figure}
\begin{center}
\includegraphics[scale=0.28]{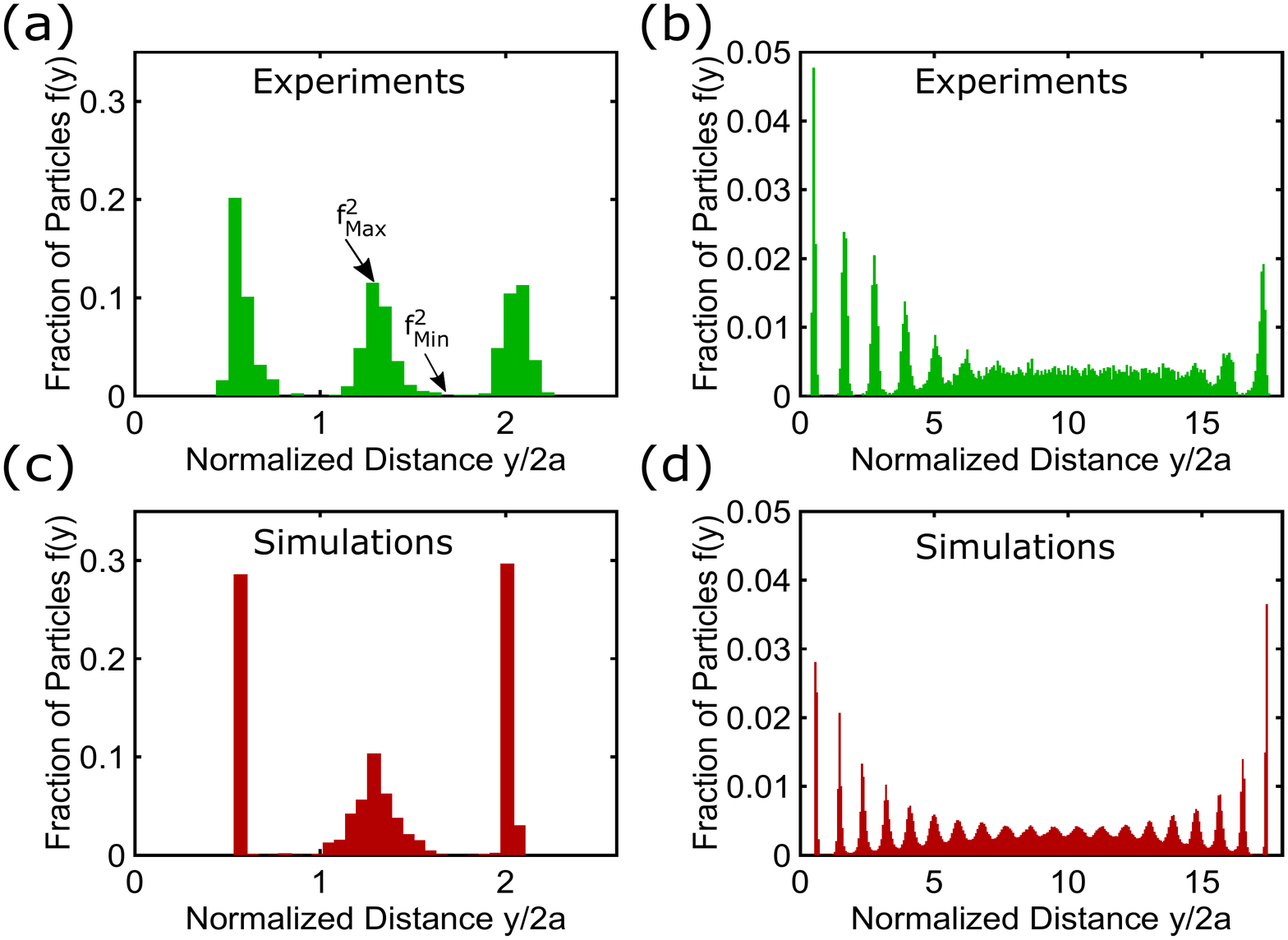}
\end{center}

\caption{Layering under confinement. Histograms of the normalized y coordinate of the particle positions from experiments (green histograms, (a) and (b)) and lubrication-repulsion simulations (red histograms, (c) and (b)) for small ((a) and (c)) and large ((b) and (d)) gaps. The bottom plate corresponds to y=0. Collectively these histograms indicate strong layering with increasing confinement.}
\label{fig:histogram}
\end{figure}

\begin{figure}
\begin{center}
\includegraphics[scale=0.55]{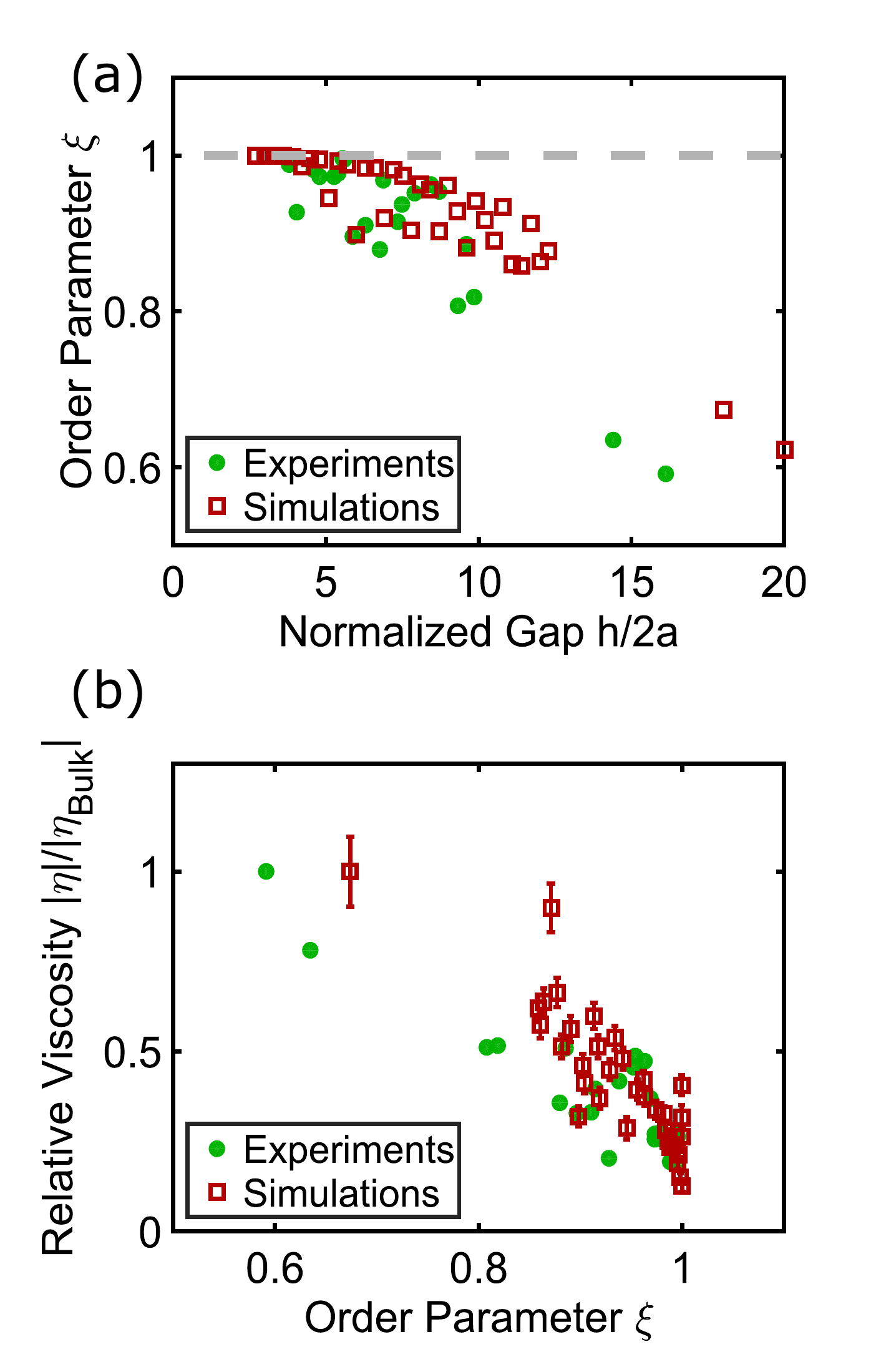}
\end{center}
\caption{Relating viscosity to the gap dependent order parameter. (a) The order parameter as a function of gap from both experiments and the lubrication-repulsion dynamics simulation. (b) The normalized viscosity as a function of the order parameter for the experiments and the lubrication-repulsion dynamics simulation. The decrease in the viscosity is well correlated with the increase in the order parameter with an R value of 0.8031.}
\label{fig:orderparameter}
\end{figure}

 The key change in the microstructure accompanying the decrease in the viscosity for 15$ >$ h/2a $>$6 is the ordering of  particles into layers parallel to the walls as the suspension is confined (Fig.~\ref{fig:apparatus}b). This layering can be seen more clearly by analyzing the confocal images obtained experimentally. We feature the particles using a standard particle featuring algorithm \cite{CrockerGrier}. Histograms of the y coordinate of the particle centers are plotted with a bin size of  0.135 microns, which is equal to the z-resolution of the microscope. Sample histograms for the small (h/2a $= 3$) and the large (h/2a $= 18$) gaps are shown in Fig.~\ref{fig:histogram}a and b respectively. In the unconfined system, there is a uniform distribution in the central region and strong peaks near the walls as is expected from the images and previous literature \cite{Maxley, Komnik, Zurita-Gotor, blanc2013microstructure}. As the gap decreases, the peaks in the histogram are more prominent, and the fraction of the particles in layers increases with strong layering visible at the smallest gap. The same analysis performed with the extracted particle positions from the lubrication-repulsion dynamics simulation as shown in Fig.~\ref{fig:histogram}c-d. The simulation results show a similar layering as the experiments. To quantify this layering, we define the order parameter: 

\begin{equation} 
\xi = 1 - \frac{1}{N}\sum_{i = 0}^{N} \frac{f^{i}_{\mathrm{Min}}}{f^{i}_{\mathrm{Max}}}
\end{equation}
where $f^{i}_{\mathrm{Min}}$ and $f^{i}_{\mathrm{Max}}$ are the heights of the i-th minima and maxima in the histogram of the y coordinates, as shown in Fig.~\ref{fig:histogram}a. The sum ranges over all the N peaks in the histogram. Thus, $\xi = 1$ for a layered sample and $\xi = 0$ for a disordered or homogeneous system. We find that with decreasing gap, $\xi$ increases (Fig.~\ref{fig:orderparameter}a) and the viscosity decreases (Fig.~\ref{fig:orderparameter}b). A linear fit to the relative viscosity versus $\xi$ data gives an R value of $0.8031$ indicating that layering is highly correlated with the decrease in the viscosity. 

\begin{figure}
\begin{center}
\includegraphics[scale=0.65]{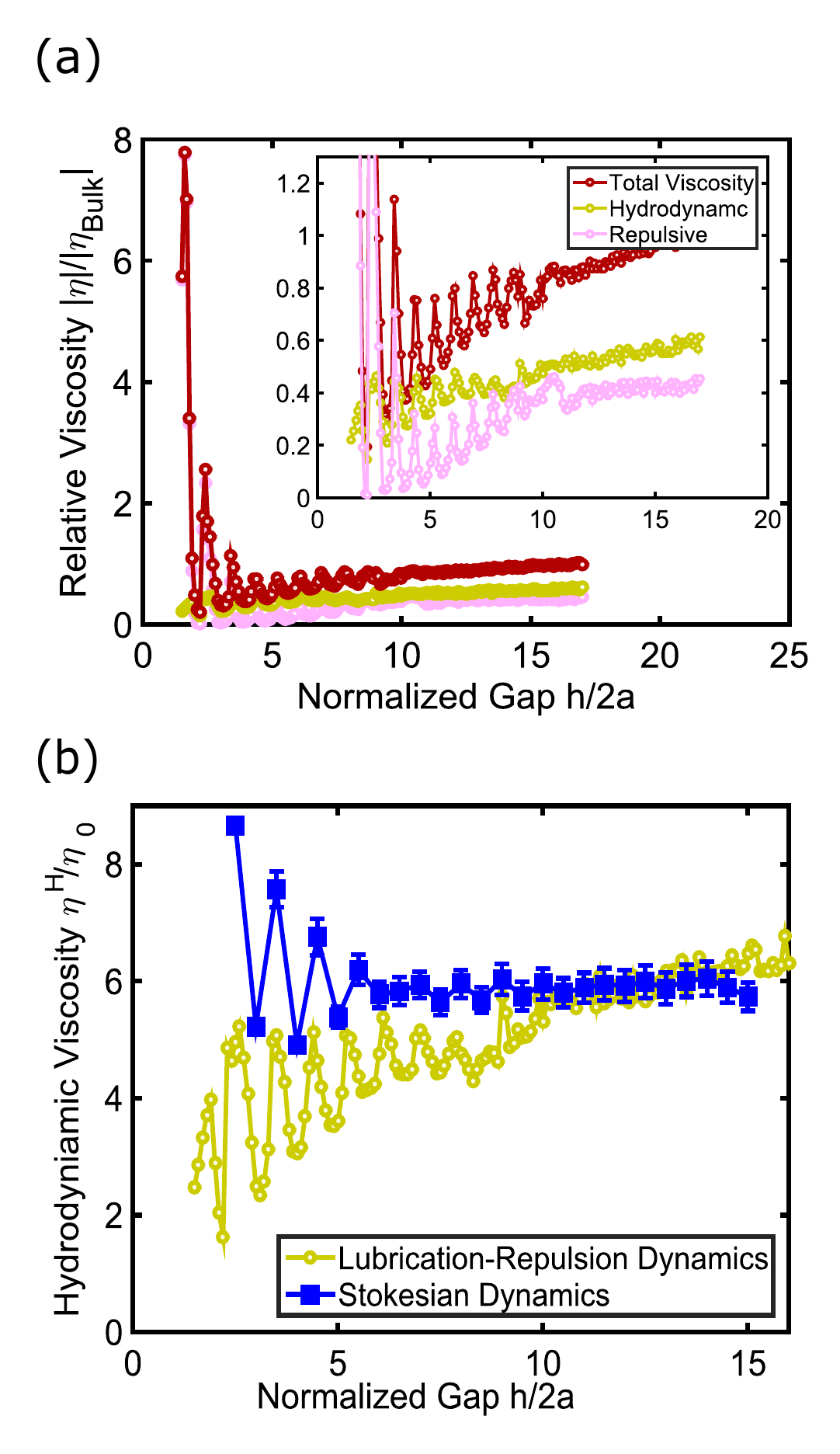}
\end{center}

\caption{Force contributions to the total stress as determined from the two simulation techniques. (a) The total (red), hydrodynamic (yellow), and short-ranged repulsion contributions (pink) to the relative viscosity of the suspension as calculated by the lubrication-repulsion dynamics simulation. The viscosities are scaled by the total bulk viscosity in all cases.  The inset shows comparable decreases in the viscosities arising from short range repulsive and hydrodynamic interactions. (b) Comparison of the hydrodynamic viscosity from the lubrication-repulsion dynamics simulation, where a lubrication approximation is used between particle pairs, and the full hydrodynamic viscosity from Stokesian dynamics. Here, the viscosity is normalized by the suspending fluid viscosity. Both simulations show similar bulk viscosities and viscosity oscillations at low gaps. The lubrication-repulsion simulation shows up to a factor of three reduction in viscosity for normalized gaps less than 10.}
\label{fig:hydrocontact}
\end{figure}

This correlation between layering and viscosity can arise from different origins. For example, layering can change the hydrodynamic viscosity by increasing the fraction of particles that follow affine trajectories and making it easier for particles to flow over one another. Layering, however, could also increase the minimum separation between the particles making the contribution from short-ranged repulsive forces smaller. To determine which mechanism dominates, we compare the hydrodynamic and short-range repulsion contributions in the lubrication-repulsion dynamics simulation (Fig.~\ref{fig:hydrocontact}a). We find a comparable decrease in the hydrodynamic and short range repulsive stresses for this regime of moderate confinement.

To determine whether the lubrication-repulsion dynamics simulation is accurately assessing the hydrodynamic stress, we compare it to that of the Stokesian dynamics simulations by plotting the hydrodynamic viscosity versus gap (Fig.~\ref{fig:hydrocontact}b). We find that the hydrodynamic interactions from lubrication-repulsion dynamics show quantitative agreement with the Stokesian dynamics simulations at large gaps but show larger decreases under further confinement, even though the Stokesian dynamics simulations also show layering under confinement. This discrepancy in the stresses calculated by the two simulation techniques suggests that there might be a long range contribution to the hydrodynamic stress at small gaps, that is neglected by the lubrication-repulsion model. We also find a difference in the microstructure formed under confinement in the two simulation techniques. The Stokesian dynamics simulations show layering but little to no alignment in the flow direction (see Appendix A), which could also contribute to the difference in the hydrodynamic viscosity at small gaps. 

%

\subsection{Buckled Phase}
\begin{figure*}
\begin{center}
\includegraphics[scale=0.6]{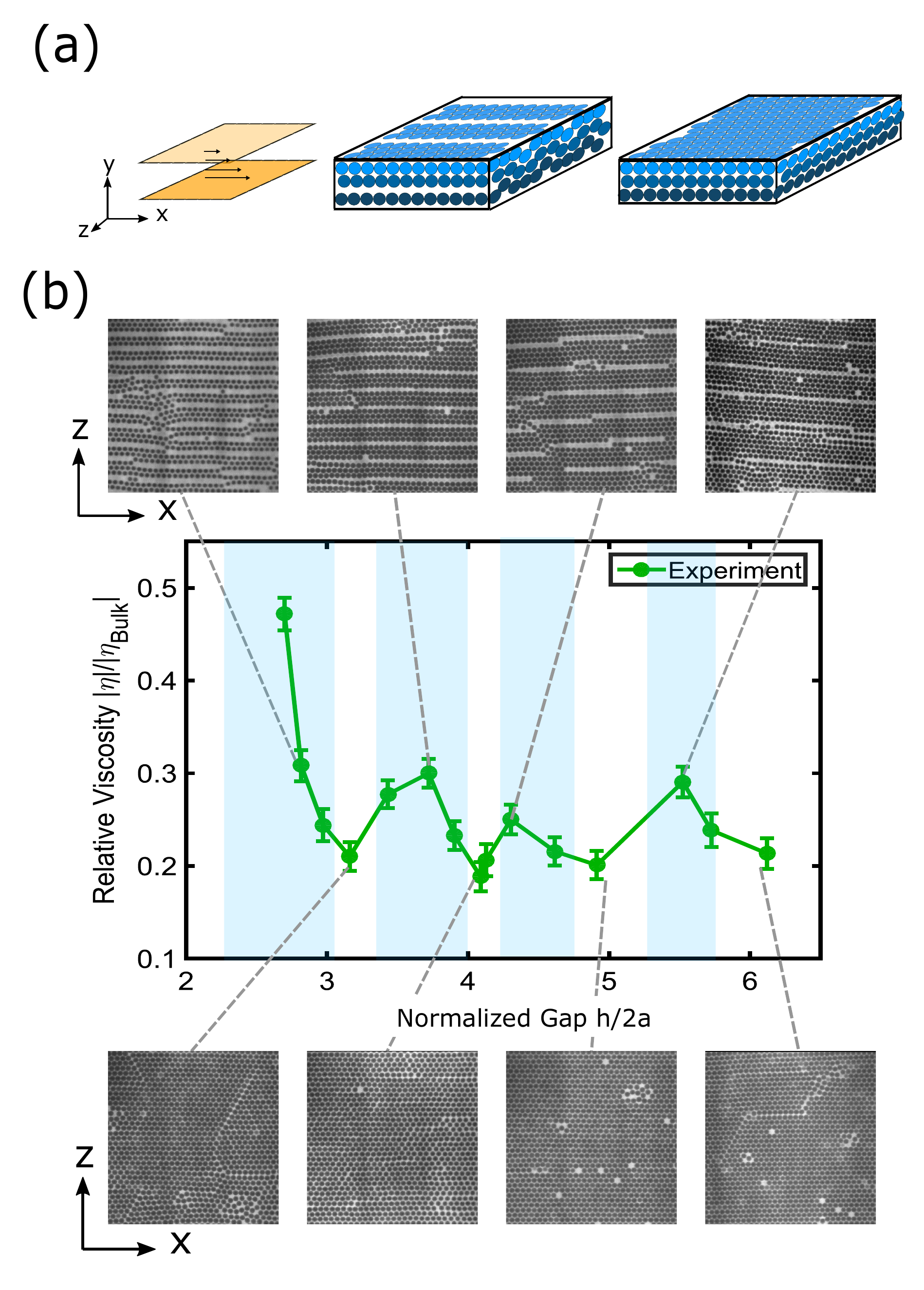}
\end{center}

\caption{Buckled phase viscosity. (a) The shear geometry and a schematic of the 3 dimensional microstructure of the buckled phase. The spheres of the same color indicate particles that move together with same velocity and displacement. (b) The magnitude of the complex viscosity as a function of the gap. The images show the x-z cross section of the microstructure at the indicated values of the gap. }
\label{fig:buckled}
\end{figure*}

As the monodispersed sample is confined further, $3 <$ h/2a $<$ 6, we observe that the viscosity fluctuates with gap. These oscillations have a length scale equal to the particle diameter (Figs.~\ref{fig:rheology},\ref{fig:hydrocontact}). Previous experimental observations of confined suspensions under shear show that when the gap is incommensurate with an integer number of particle layers, the system forms a buckled phase under shear \cite{Cohen}, where the particle layers fold out of plane (Fig.~\ref{fig:buckled}a). Our experiment and simulation data suggest that such phases may be responsible for the viscosity oscillations. To test this hypothesis, we image the sheared suspension structure over this range of gap. We find that when the gap is incommensurate with an integer number of layers, the particles form a buckled phase \cite{Cohen} and the relative viscosity is higher (Fig.~\ref{fig:buckled}b). The magnitude of these oscillations is seen to increase with smaller gap.

These oscillations in the viscosity seen in the experiments can be reproduced using the lubrication-repulsion dynamics simulation (Fig.~\ref{fig:hydrocontact}a). In the simulations, however, these oscillations have a much larger amplitude and less well formed structure (Appendix~A). Separating the short range repulsive and hydrodynamic contributions, we find that the fluctuations in the viscosity arise from both forces, with short range repulsion playing a larger role at smaller gaps (Fig.~\ref{fig:hydrocontact}a). Comparing with Stokesian dynamics simulations (Fig.~\ref{fig:hydrocontact}b), we find good agreement with the amplitude and gap dependence of the oscillations. We do however find more significant decrease in the hydrodynamic contribution to the viscosity in the lubrication-repulsion model. Collectively these data demonstrate that confinement induced microstructure and geometric incommensurability strongly affect the hydrodynamic and short range repulsive forces giving rise to the suspension viscosity.

\subsection{Extreme Confinement}
Finally, at extremely small gaps (h/2a $<$ 3), we find that the viscosity amplitude sharply increases (Figs.~\ref{fig:rheology},\ref{fig:hydrocontact}, ~\ref{fig:buckled}b). A viscosity increase is also observed at the same normalized gap for the lubrication-repulsion dynamics simulations. However the increase in the simulations is much larger (Fig.~\ref{fig:hydrocontact}a), which may be due to the even smaller gaps reached in simulations. On separating the hydrodynamic and the short range repulsive contributions to the stress, we find a small increase in the hydrodynamic stress in the lubrication-repulsion simulations. This increase is comparable to the increase seen in the Stokesian dynamics simulations. Collectively, these data indicate the large increase in the viscosity at extreme confinement primarily arises from the short range repulsive forces. 



\section{Tuning the suspension rheology under confinement}

\begin{figure}
\begin{center}
\includegraphics[scale=0.65]{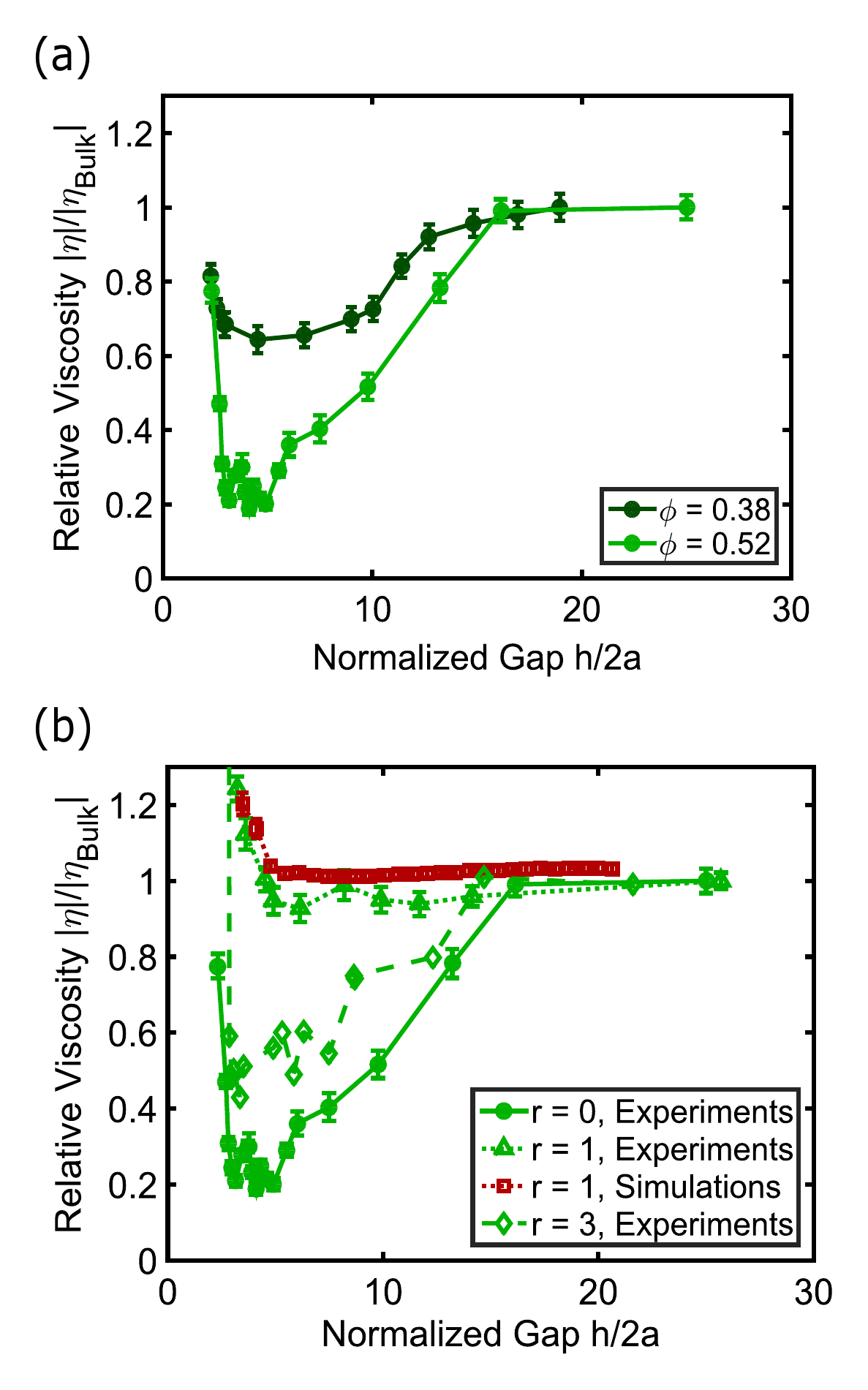}
\end{center}

\caption{Tuning the viscosity under confinement (a) The decrease in the viscosity under confinement for 2 different volume fractions. (b) The viscosity of a bidisperse system at three different number ratios of small to large particles, r = 0, 1, and 3. Also shown are the lubrication-repulsion simulation results for r = 1.}
\label{fig:bidispersed}
\end{figure}

The structural dependence of the hydrodynamic and short ranged repulsive stresses indicates that volume fraction and polydispersity could be used as additional knobs for tuning the suspension response under increasing confinement. For example, decreasing the volume fraction results in a suspension that is significantly less layered under confinement. Thus, such suspensions should exhibit a smaller reduction in the viscosity with gap. To test this prediction we compare the viscosity versus gap measurements for suspensions with volume fraction $\phi=0.38$ and $0.52$ (Fig.~\ref{fig:bidispersed}a). We find that as anticipated the lower volume fraction suspension shows significantly less decrease in viscosity under confinement. At these volume fractions the suspension microstructure never forms full layers. As such, the viscosity oscillations arising from the buckled phases are also absent (Appendix~B). We note that denser suspensions that are crystalline even in bulk would be layered at all gaps. Thus, they are expected to show little to no decrease in viscosity with moderate confinement. We would, however, expect such systems to exhibit buckled phases and the corresponding viscosity fluctuations. Finally, in all cases we expect to observe the sharp increase in viscosity under extreme confinement.   

Polydispersity can also be used to inhibit layer formation. Thus, we predict that the decrease of viscosity due to confinement induced ordering would diminish with increasing polydispersity. To test this prediction we measure the relative viscosity versus normalized gap for bidisperse suspensions with different degrees of polydispersity. The suspensions are comprised of two different particles with incommensurate diameters, 2 $\mu m$ and 1.3 $\mu m$. We control the extent to which the suspensions can layer by changing the number ratio of the small to the big particles $r$. For example, using $r = 1$ we can completely suppress the layering in the systems and we observe a constant viscosity down to very small gaps (Fig.~\ref{fig:bidispersed}b green and red lines). In contrast using $r = 3$ we observe some layering in the suspension and hence a small but significant decrease in the viscosity is observed (Fig.~\ref{fig:bidispersed}b diamonds symbols). These experiments demonstrate that the suspension microstructure is a powerful tool that can be utilized to tune the confined suspension viscosity.

\section{Discussion}

\subsection{Viscosity fluctuations arising from buckled phase microstructure}
The measured increase in the viscosity when the suspension forms a buckled phase contradicts the model put forward previously by Cohen et al. \cite{Cohen}. This prior work suggested that the amplitude of the shear stresses is proportional to the shear dependent osmotic pressure \cite{bergenholtz2002non} in the shear zone. Since the shear dependent osmotic pressure must balance the constant osmotic pressure in the reservoir, it was predicted that the effective viscosity of the suspension must also be constant for all gaps.  While we find that the viscosity increase in the buckled structures observed in our experiments is moderate, indicating the osmotic pressure may set the overall scale of the viscosity, our measurements suggest that the coupling between the viscosity and the osmotic pressure is more complex.

More specifically, the complication arises from the fact that details of the suspension structure can alter the normal stresses generated by a given shear flow. For example, when the suspension forms a buckled phase, a normal stress in the gradient direction pushing a particle sitting below its neighbors will be redirected laterally through the short ranged repulsion between the particles. Thus, the degree to which forces in different directions are coupled may vary substantially for different structures and requires further investigation.

In principle, numerical simulations that impose a buckled phase structure parametrized by the extent of buckling could be used to fully elucidate the origin for the viscosity increase. For example, such studies could be used to determine the coupling between the reservoir osmotic pressure and the shear stresses generated by different degrees of buckling. In addition, for a given structure and flow, the hydrodynamic and short range repulsion contributions to the shear stresses could be determined. Such a study would also allow for distinguishing whether effective surface area between layers or distance between layers dominate the increase in hydrodynamic contributions in the buckled phase.

\subsection{Increase in viscosity under extreme confinement}

\begin{figure}
\begin{center}
\includegraphics[scale=0.65 ]{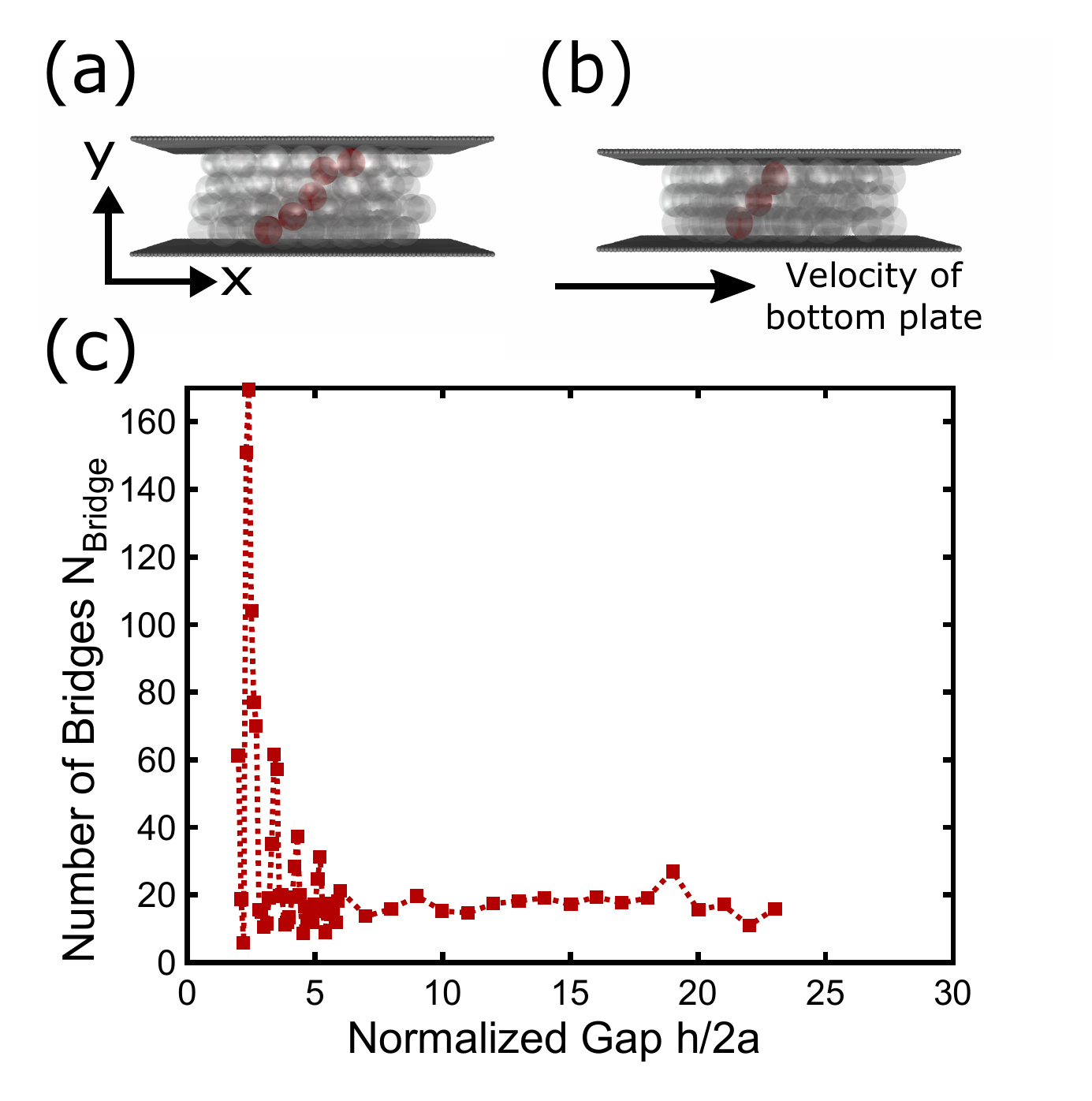}
\end{center}

\caption{Bridge formation under confinement. (a) and (b) show an example of a bridge seen in the simulations at gaps of 3.5 and 2.5 particle diameters respectively. (c) The number of system spanning bridges as a function of gap. We find a sharp increase in the number of bridges corresponding to the increased viscosity at extreme confinement.}
\label{fig:ultraconfined}
\end{figure}

At gaps h/2a $<$ 3, the suspension viscosity increases dramatically. The simulation data indicate that short ranged repulsion provides the dominant contribution to this increase. Motivated by the formation of force chains in granular systems, we track the increase in the number of ``bridges'' that span the system between the two walls. Here, a bridge refers to an uninterrupted chain of particles whose interactions are dominated by short ranged repulsive forces. We use the simulation data from the lubrication-repulsion dynamics simulations to plot the number of system spanning bridges for the mono-dispersed system versus normalized gap. We find that the number of bridges increases sharply at a normalized gap of three. These results indicate that while short range repulsion forces contribute at all gaps, the sharp increase in the viscosity at extreme confinement arises from the increase in the bridges between the upper and lower walls (Fig.~\ref{fig:ultraconfined}).  

\subsection{Comparisons to atomic and granular systems} 

Many of the changes in the viscosity of a colloidal suspension under confinement can be compared to those observed in atomic and granular systems. In atomic systems for example, it has been shown that when water is confined such that the gap size is comparable to a few times that of the molecule, its viscosity increases by several orders of magnitude \cite{ Dhinojwala, Israelachivili1}. The similar increase seen in colloidal suspensions under extreme confinement is consistent with the idea that formation of short load bearing bridges between the confining surfaces may be the underlying cause of the viscosity increase in atomic fluids. This mechanism also explains why the viscosity increase only occurs at very small gaps: without friction or some other mechanism that prevents lateral slipping between particles or atoms, it is difficult to support long force chains between the plates. 

Simulations and experiments in atomic systems have also shown that extreme confinement can induce structural ordering that depends sensitively on the gap \cite{Israelachivili2, Klien1}. In particular, it has been shown that incommensurability of the gap with the atom size results in oscillations in the viscosity \cite{Neek-Amal, Israelachivili3}. These viscosity oscillations closely resemble those seen in colloidal suspensions when the gap in incommensurate with an integer number of particle layers. It would be interesting to determine whether structures similar to the observed colloidal buckled phases also arise in atomic systems.

Monodispersed granular suspensions also display trends similar to colloidal systems under confinement \cite{PeylaExp, Maxley, Brown}. Granular suspensions show an increase in the viscosity at gaps smaller than three particle diameters. While some papers suggest that this increase is due to hydrodynamic interactions between the particles and the boundaries \cite{PeylaSim}, others suggest that friction may play a role in this increase in viscosity \cite{Brown}. Our current results showing the larger contribution from the short range repulsive forces suggests that friction may be the dominant factor in the increase in viscosity in granular suspensions. 

At moderate concentrations ($\phi = 0.2 - 0.4$), simulations show that the viscosity initially decreases when a granular suspension is confined to gaps less than $15$ particle diameters before increasing when the gap is less than 3 particle diameters \cite{PeylaSim}. This decrease in the viscosity is very similar to the decrease in the viscosity seen in colloidal suspensions and could also be the result of the layering due to the presence of boundaries. 

At higher volume fractions ($\phi = 0.58$), granular suspensions no longer show this decrease in the viscosity, and the viscosity remains constant until the gap is smaller than $\sim 7 $ particle diameters. In light of our results, it may be the case that the monodisperse granular suspension has already ordered during confinement. For example, it has been shown via simulations and experiments that granular systems layer parallel to the wall under confinement \cite{Fornari, Brown}. Such ordering would rule out the decrease in viscosity due to the layering mechanism that is observed in the present study. Moreover, such ordering would still preserve the viscosity oscillations for gaps below $\sim 7 $ particle diameters \cite{Brown, Maxley}.

The results of our experiments also show similarities with simulations of confined suspensions at higher Reynolds number \cite{Fornari}. Those simulations show a similar decrease and fluctuations in the viscosity even at volume fractions as low as $\phi = 0.3$. They also demonstrate layering in the suspension under confinement, and show an increase in viscosity at gaps incommensurate with the particle diameter. Such results hint that inertia could lead to additional mechanisms that enhance layer formation in commensurate gaps and give rise to oscillations in the viscosity. 

\section{Conclusions}
Our experiments and simulations show that the structures that arise due to confinement play an essential role in setting the balance of forces that determine the viscosity of the suspension. For a monodispersed sample with high volume fraction ($\phi = 0.52$), we find that the viscosity decreases at moderate degrees of confinement because of the layering that arises due to the presence of the walls. This layering gives rise to comparable decreases in the hydrodynamic and short ranged repulsive forces, both of which contribute significantly to the viscosity. Further, when the gap is less than 6 times the particle diameter, the formation of a buckled structure increases the viscosity for gaps that are incommensurate with particle layers. These structural variations again give rise to comparable changes in the hydrodynamic and short ranged repulsive forces. Finally, under extreme confinement, when h/2a $<$ 3, the viscosity sharply increases due to particle bridging between the plates. This increase is dominated by the short ranged repulsion forces between the particles.

This complex relationship between the viscosity, microstructure and confinement enables us to tune the suspension rheology by altering the gap, volume fractions, and polydispersity of the suspension. In addition, the formation of anisotropic structures such as the buckled phase, which is aligned along the shear direction, suggests the suspension viscosity may be anisotropic. Finally, the study presented here has only explored the effect of confinement at intermediate Pe numbers. The effects of confinement at very low Pe numbers (Brownian regime) and very large Pe numbers (shear thickening regime) remain open for future investigations.  

\subsection{Acknowledgments}
 The authors would like to thank the Cohen lab members and Jin Sun for helpful discussions. M.R., I.C., N.Y.C.L., and B.D.L. were supported by NSF CBET-PMP Award No. 1232666 and continued support from NSF CBET-PMP Award No. 1509308.C.N. was supported by the Maudslay-Butler Research Fellowship at Pembroke College, Cambridge. A.F. and J.W.S. were supported by MIT Energy Initiative Shell Seed Fund. 

\subsection{Author contributions}
M.R., N.Y.C.L., B.D.L., and I.C. were responsible for the colloid measurements. C.N. was responsible for the lubrication-repulsion simulations. A.F. and J.W.S. were responsible for the Stokesian dynamics simulations. All authors contributed to the writing of the manuscript.   

\section{Appendix A}
\begin{figure}
\begin{center}
\includegraphics[scale=0.65 ]{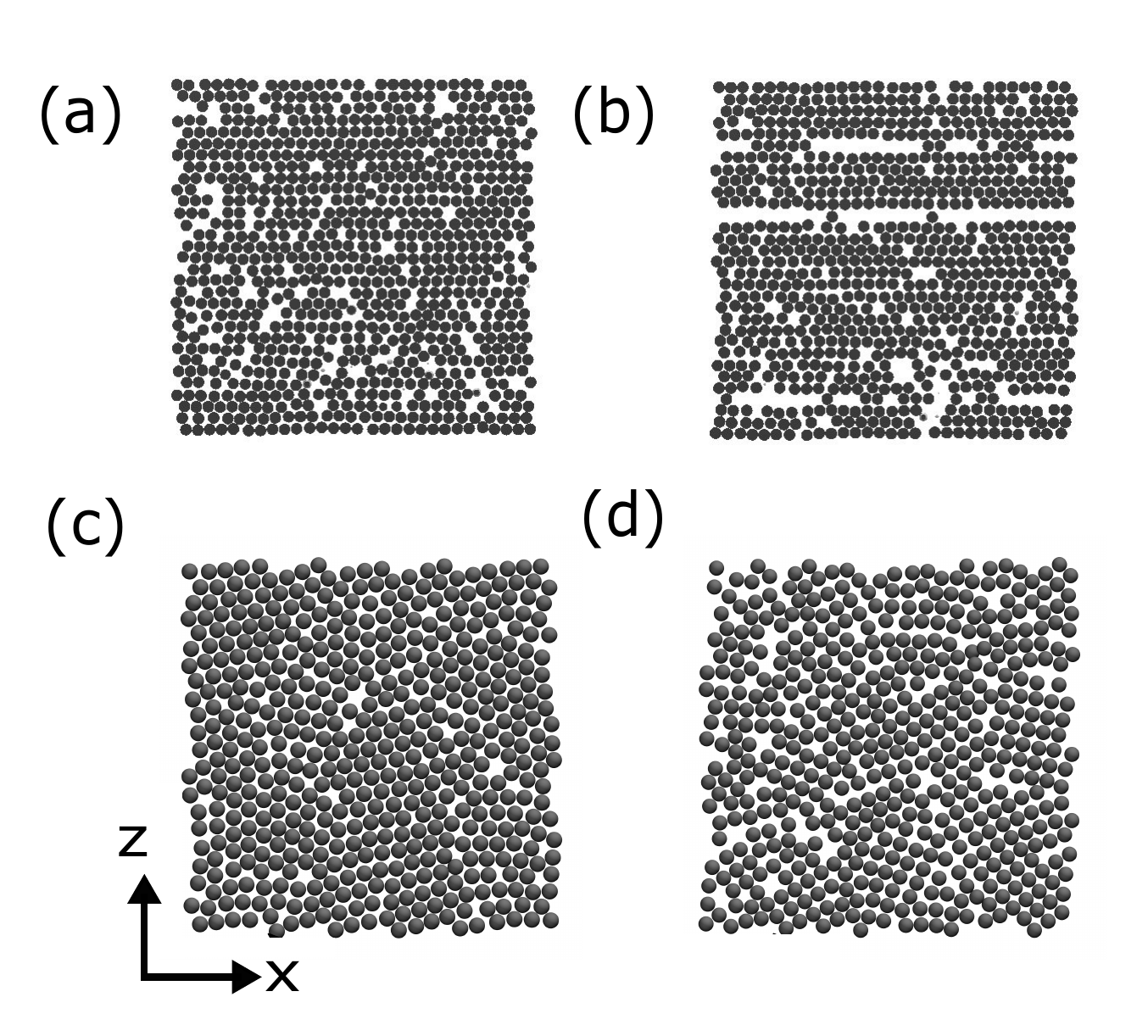}
\end{center}

\caption{The suspension microstructure in the shear  vorticity plane from simulations. (a) and (b) show the structure formed in the  lubrication-repulsion simulations at gaps 3.5 and 3.9 particle diameters respectively. (c) and (d) show the structure formed in the Stokesian dynamics simulations at gaps 3.5 and 4 particle diameters respectively.}
\label{fig:buckledsim}
\end{figure}

We show here the structures formed in the shear vorticity plane in the simulations at high volume fractions ($ \phi = 0.52$). Figs.~\ref{fig:buckledsim} (a) and (b) shows the microstructure at gaps 3.5 and 3.9 particle diameters from the lubrication-repulsion simulations. At incommensurate gaps, we see the stripes characteristic of the buckled phase (Fig.~\ref{fig:buckledsim}a). Comparing with Fig.~\ref{fig:buckled}b, we see that the experimental images are more periodic, which may be the result of larger system size, Brownian motion as well as smoother walls in the experiments. Figs.~\ref{fig:buckledsim} (c) and (d) show the microstructure at gaps 3.5 and 4 particle diameters from the Stokesian Dynamics simulations. We see layering during confinement but in contrast to the lubrication repulsion simulations and experiments, the particles are not aligned along the flow direction. At gaps incommensurate with the particle diameter, the ordered domains indicative of layering formed at commensurate gaps are broken up. However, there is no evidence of the formation of the stripes seen in the buckled phase in the Stokesian Dynamics simulations (Fig.~\ref{fig:buckledsim}d). These structural differences suggest that control of boundary conditions along the vorticity direction may be a key factor in simulating aligning dispersions as flow alignment is impeded when the simulation cell dimensions are incommensurate with an integer number of flow aligned particles in the Stokesian Dynamics simulations. The difference in microstructure between the experiments and the simulations very likely contributes to the quantitative difference in the suspension viscosity (Fig.~\ref{fig:rheology}, Fig.~\ref{fig:hydrocontact}b).

\section{Appendix B}
\begin{figure}
\begin{center}
\includegraphics[scale=0.4 ]{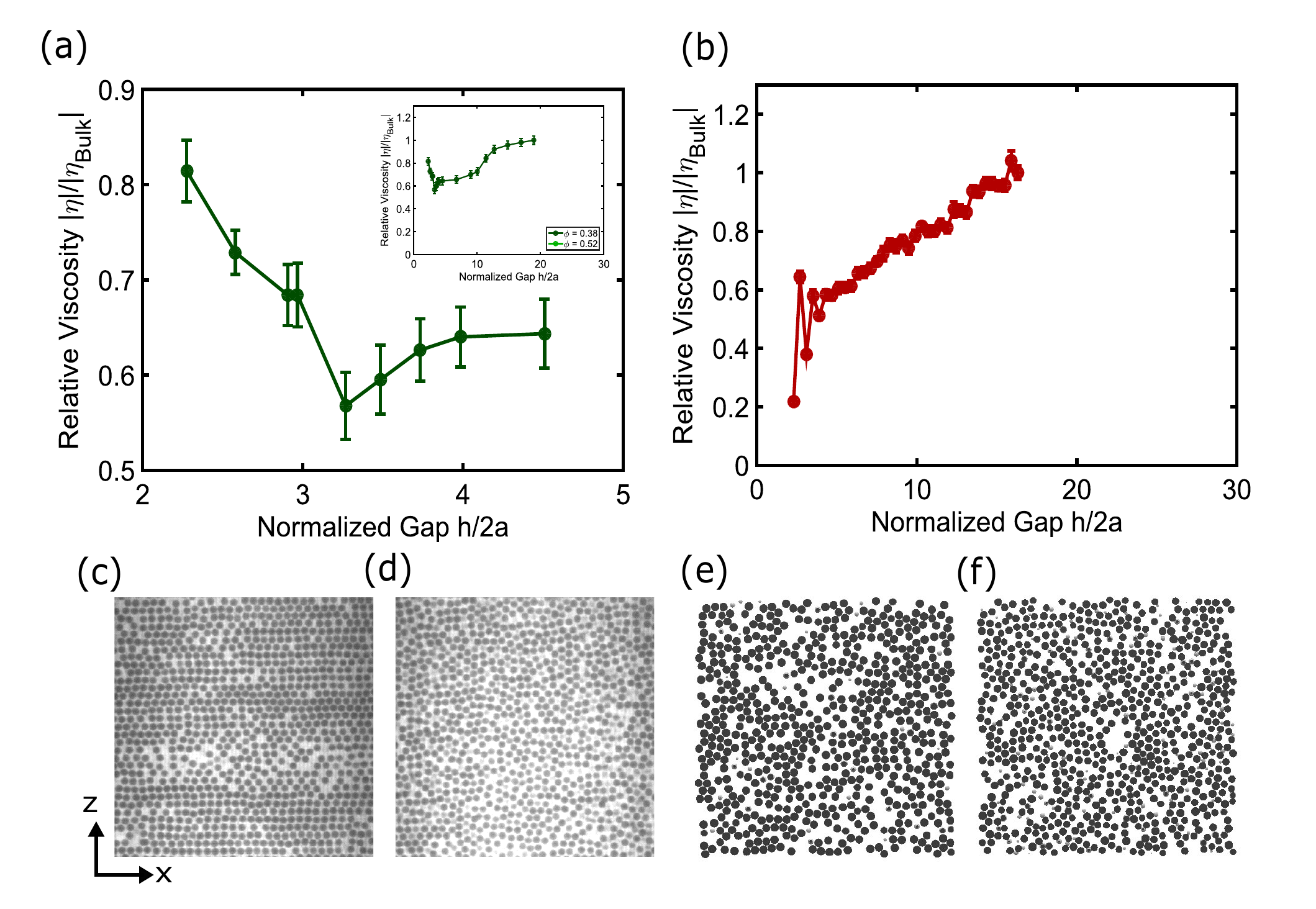}
\end{center}

\caption{Rheology and microstructure of a low volume fraction ($\phi = 0.38$) monodisperse suspension.(a) The experimentally observed variations in the viscosity between gaps 2.5 to 4 particle diameters. We see no measurable oscillations due to incommensurability. The inset shows the  viscosity variation over the full range of gaps. (b) The viscosity of the suspension measured by the lubrication-repulsion simulations. (c), (d), (e) and (f) show the microstructure from experiments ((c) and (d)) and simulations ((e) and (f)). (c) and (e) are at a gap of 3.5 particle diameters and (d) and (f) at a gap of 3.9 particle diameters. Collectively, the structural data demonstrates the absence of layering and formation of the buckled structure in the sample.}
\label{fig:lowvolfrac}
\end{figure}

We discuss here in more detail the rheological and structural trends seen in suspension of low volume fraction ($\phi = 0.38$). At this volume fraction, we expect the layering to be decreased and no buckled structures to be formed. In agreement with our expectations, the results of the experiments show a smaller decrease in the viscosity (Fig.~\ref{fig:lowvolfrac}a(inset)). Moreover, we see no measurable variations in the viscosity due to incommensurability of gap with the particle diameter (Fig.~\ref{fig:lowvolfrac}a). The microstructure also displays less layering (Fig.~\ref{fig:lowvolfrac}c and d) and no buckled phase is formed when the gap is incommensurate with the particle diameter (Fig.~\ref{fig:lowvolfrac}c). The lubrication-repulsion dynamics is expected to be less accurate at $\phi < 0.4$. Nevertheless, we run the simulations at $\phi = 0.38$. We observe a decrease in viscosity comparable with that seen in experiments. However, fluctuations are observed in the lubrication-repulsion simulations, even though there is no indication of the formation of the buckled phase (Fig.~\ref{fig:lowvolfrac}e). 

    \bibliographystyle{ieeetr}
\bibliography{ConfinementBib}

\end{document}